\documentclass{article}

\usepackage[dvipsnames]{xcolor}
\usepackage{sigbovik2023_conference,times}
\PassOptionsToPackage{hyphens}{url}
\usepackage[hidelinks]{hyperref}
\usepackage{url}
\usepackage{CJKutf8}
\usepackage{soul}
\usepackage{dirtytalk}
\usepackage{epigraph}
\usepackage[labelformat=simple]{subcaption}
\usepackage{xspace}
\usepackage{pifont}
\usepackage{amssymb}
\usepackage{booktabs}
\usepackage{algorithm}
\usepackage{algpseudocode}
\usepackage{frcursive}
\usepackage{tabularx}
\newcolumntype{C}{>{\centering\arraybackslash}X}
\usepackage{multirow}
\usepackage{listings}
\usepackage{amsmath}
\usepackage{graphicx}

\usepackage{wrapfig}

\usepackage{tikz}
\usetikzlibrary{arrows.meta}

\algdef{SE}[SUBALG]{Indent}{EndIndent}{}{\algorithmicend\ }%
\algtext*{Indent}
\algtext*{EndIndent}

\usepackage{pgfplots}
\pgfplotsset{width=15cm, height=7cm,compat=1.8,grid style={gray,opacity=0.25}}

\definecolor{codegreen}{rgb}{0,0.6,0}
\definecolor{codegray}{rgb}{0.5,0.5,0.5}
\definecolor{codepurple}{rgb}{0.58,0,0.82}
\definecolor{backcolour}{rgb}{0.95,0.95,0.92}

\lstdefinestyle{pystyle}{
    backgroundcolor=\color{backcolour},   
    commentstyle=\color{codegray},
    keywordstyle=\color{codegreen},
    numberstyle=\tiny\color{codegray},
    stringstyle=\color{codepurple},
    basicstyle=\ttfamily\footnotesize,
    breakatwhitespace=false,         
    breaklines=true,                 
    captionpos=b,                    
    keepspaces=true,                 
    numbers=left,                    
    numbersep=5pt,                  
    showspaces=false,                
    showstringspaces=false,
    showtabs=false,                  
    tabsize=2
}

\usepackage[capitalize]{cleveref}
\crefname{section}{Sec.}{Secs.}
\Crefname{section}{Section}{Sections}
\Crefname{table}{Table}{Tables}
\crefname{table}{Tab.}{Tabs.}

\title{
Large Language Models are Few-Shot \\Publication Scoopers
}

\author{Samuel Albanie,
Liliane Momeni,
Jo\~{a}o F. Henriques\\
Shelfanger, United Kingdom
}

\iclrfinalcopy 

\begin{document}
\maketitle

\begin{abstract}

Driven by recent advances AI, we passengers are entering a golden age of scientific discovery.
But golden for whom?
Confronting our insecurity that others may beat us to the most acclaimed breakthroughs of the era, we propose a novel solution to the long-standing personal credit assignment problem to ensure that it is golden for us.
At the heart of our approach is a \textit{pip-to-the-post} algorithm that assures adulatory Wikipedia pages without incurring the substantial capital and career risks of pursuing high impact science with conventional research methodologies.
By leveraging the meta trend of leveraging large language models for everything, we demonstrate the unparalleled potential of our algorithm to scoop groundbreaking findings with the insouciance of a seasoned researcher at a dessert buffet.

\end{abstract}

\epigraph{If I have seen farther it is by standing on the shoulders of giants. And then stealing their binoculars.}{Isaac Newton, \emph{\#daily-research-hacks}}

\section{Introduction} \label{sec:intro}

When Isaac Newton raced ahead of Robert Hooke and defied the Royal Society's Social Media Ban to promote his inverse-square law of gravity pre-print in 1686,
he exemplified the glorious pursuit of scientific priority%
\footnote{This was far from Newton's only scientific priority fracas.
Asked what he thought of Leibniz' work,  Newton quipped ``derivative'', before laughing so hard that infinitesimal tears ran down his cheeks.}
that has long galvanised boffins the world over.%
\footnote{
Newton's \textit{Principia} was financed by Halley, who'd discussed the problem with both Newton and Hooke.
The Royal Society had planned to fund Newton's publication, but they had entirely exhausted their book budget on \textit{De Historia Piscium} (Of the History of Fish), by Francis Willughby, a scholarly work that surprisingly failed to achieve best-seller status.
}

Unfortunately, the unrelenting pursuit of personal credit assignment is an activity in decline.
Few modern scientific feuds match the intensity of the late 16th century public \textit{Priorit\"atsstreit}\footnote{A ``priority dispute''. We've used German to remind the reader that this is serious business.} between astronomers Tycho and Ursus over credit for the geoheliocentric model (a spat that involved, \textit{inter alia}, dramatic midnight raids on bedrooms to retrieve allegedly stolen diagrams from trouser pockets~\citep{worrall1985birth}).

Instead, fields such as Machine Learning, which could long be relied upon to generate such drama, have degenerated into a head-to-head showdown with Particle Physics in a quest to show which is more of a ``team sport'' through feats of collaboration%
\footnote{
At the time of writing, High Energy Physics maintains a comfortable lead, with a 5,154 author paper estimating the size of the Higgs Boson~\citep{aad2015combined}. 
}.
Indeed, fuelled by a seemingly inexhaustible supply of memes, technical prowess
and \textit{esprit de corps},
distributed open-source collectives now represent a major contributor of high-impact breakthroughs.
In tandem, well-funded technology firms have gathered their researchers into ever larger familial structures and task forces.%
\footnote{This excludes the CFO, who instead nervously increments variables on the communal \texttt{slurm.conf}.}

This all sounds wonderfully warm and fuzzy, but let us consider its consequence for those of us with the onerous time commitments of hourly checking our Google Scholar profile and Twitter follower count, prohibiting effective participation in such teams.
Modern reviewers, unaware of whether they are reviewing a submission from three authors or thirty-three, have high expectations.
Standards have been raised.
The sad result is that meaningful contributions in the era of big-discord-science have become terribly hard work.

\begin{wrapfigure}{r}{0.25\textwidth}
    \centering
    \vspace{-0.3cm}
    \includegraphics[width=0.25\textwidth]{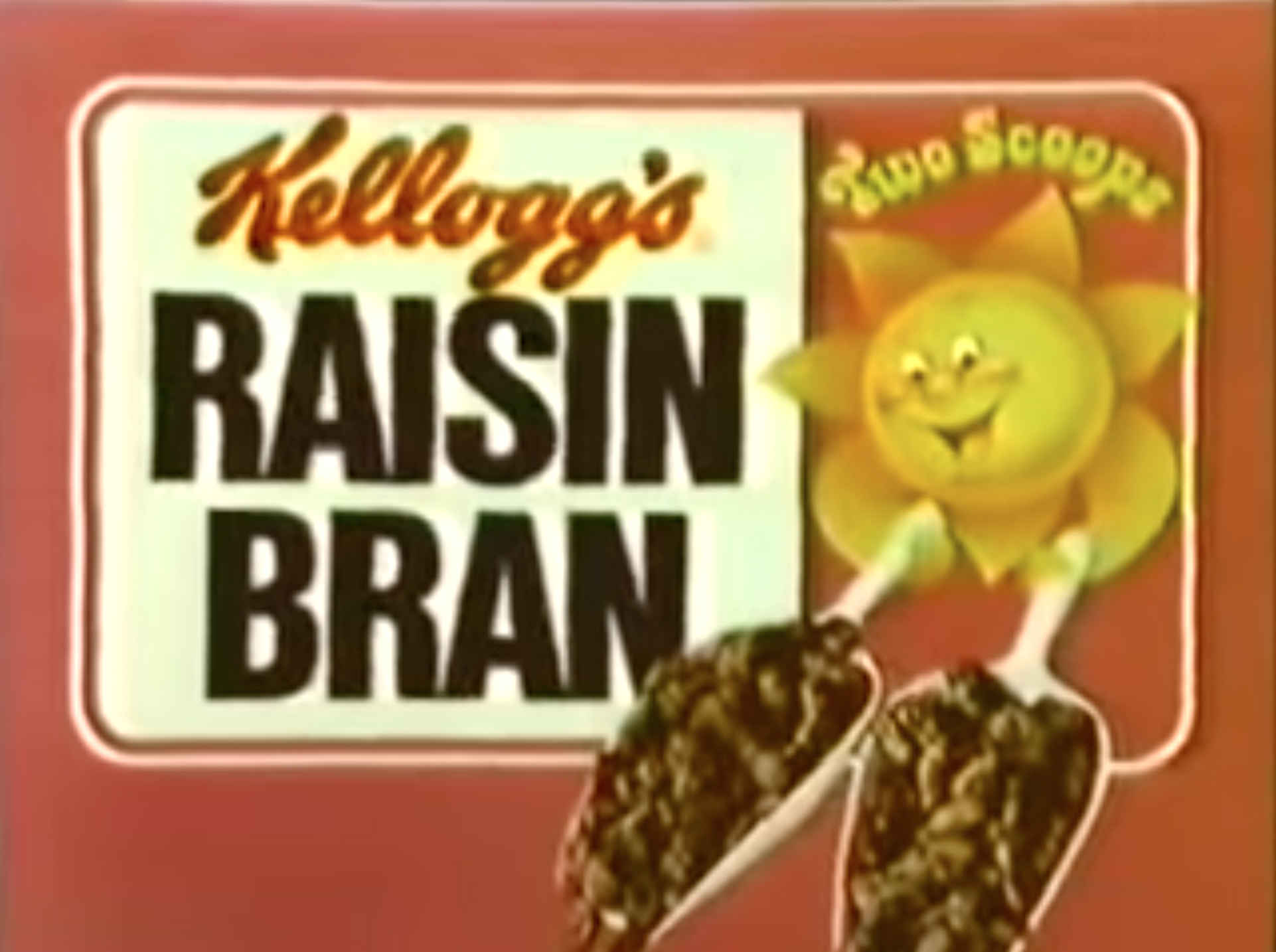}
    \caption{\textbf{Award certificate presented at CVPR 1983}. Entitling authors obtaining two scoops to a deliciously fibrous breakfast.}
    \label{fig:2scoops}
\end{wrapfigure}

Even if we were to develop some self control and find time to join these teams, there is a second problem.
The \textit{whole point} of doing science is to achieve personal glory while strongly signalling that we are not motivated by a desire for personal glory.
It is entirely natural to harbour a healthy clandestine lust for prizes, international fame and a lifetime supply of Cheerios from an adoring sponsor.
But once we shackle ourselves to a high-performing team, who receives the credit?
It would be simply awful to contribute a breakthrough and then be forced to share the Cheerios.
After all, as wisely noted by the Nobel committee, the maximum number of people that can possibly discover something interesting is three.

In this work, we propose the use of \textit{scooping}---the act of publishing an important result before others who pursue a similar agenda---as a novel, efficient and practical solution to the Cheerios problem. A baseline of ``Two Scoops'' has long been considered sufficient for sponsorship by Kellogg's Raisin Bran (\cref{fig:2scoops}), but we crave tastier cereal and unbounded scoops. %
Thus, while scooping to date has been a largely passive affair, we draw inspiration from Ursus' purported plagiarism of Tycho and develop an \textit{active} scientific scooping framework as a basis for our solution.

We make three contributions.
First, we formalise the Cheerios problem.
Second, we advance arguments for the increased algorithmic and financial efficiency of proactive scooping over the existing (largely-passive) scooping paradigm for resolving this challenging breakfast dilemma.
Third, we demonstrate practical few-shot active scooping by leveraging a recent increment in the absolutely concerning series $\{\textrm{GPT-}n: n \in \mathbb{N}\}$, a 7-day free trial premium Overleaf subscription and 104 Twitter puppet accounts to scoop multiple high-impact publications on the topic of robust flower breed classification.

\epigraph{I certainly should be vexed if any one were to publish my doctrines before me. I want me those Cheerios.}{Charles Darwin, 1856}

\section{Related Anti-Teamwork}

\textbf{Scientific Priority.}
Seminal work by~\cite{merton1957priorities} established the key role of scientific priority as a reward signal to encourage originality (mildly tempered by a respectable emphasis on humility%
\footnote{In addition to humility, certain fields, such as mathematics, also encourage \textit{understatement}. 
This likely stems from a healthy fear of exclamation marks.
There are few things more explosive than a misplaced factorial.
}).
\cite{Kuhn1962HistoricalSO} observed that it was often simply impossible to assign scientific priority to an individual when a ``discovery'' does not constitute an isolated event.
That shouldn't stop us trying to both assign and claim priority.
Differently, from prior work that has sought to understand the phenomenon of scientific priority, we focus on the application of Large Language Models to its accrual.

\textbf{Scooping.}
The rush to preempt a competitor has long engaged the titans of science. 
Prior to the inconvenient loss of his head, Lavoisier scooped his rival Priestly to claim the discovery of Oxygen.
Watson and Crick openly discuss their strenuous efforts in 1953 to beat Wilkins and Franklin to the DNA structure~\citep{watson1968double}.
Even the gentle Darwin was spurred into action in 1858 by learning that Wallace had crafted a similar theory and might publish before him.
While these researchers limited themselves to scoops that fall within their expertise, we propose to use Large Language Models to broaden the scooping scope to fields that we are entirely ignorant of (watch out, petrologists).

\textbf{The value of moral flexibility.}
Ever since \cite{feyerabend1975} determined Science to be a lawless land where ``anything goes'', methods such as $n$-Dimensional Polytope Schemes~\citep{fouhey2013n} and Deep Industrial Espionage~\citep{albanie2019deep} have rigorously demonstrated the remunerative benefits of a flexible moral attitude.
We purloin the underhand theme of their work, but eschew monetary gain and instead dedicate ourselves to the pursuit of the nobler prize of achieving stellar reputations. %

\textbf{Few-shot Learning with Large Language Models.}
Let's face it, large language models can few-shot everything now. 
It's more than a little scary.
They can sing. 
They can dance.
They can scoop.

\epigraph{The best way to predict the future is to scoop it.}{Alan Kay}

\section{Method}

\textbf{The Cheerios problem.}
As humanity peeks nervously out from under her comfort blanket, she sees the intimidating dance of bedroom wall shadows cast by problems that must be confronted.
Failed AI alignment, engineered pandemics and nuclear end-games.
Food insecurity, global poverty, military conflicts and climate destabilisation. 
Those white plastic sporks that snap on pasta that exhibits the slightest hint of \textit{al dente}~\citep{ord2020precipice}.

To reach the safety of the morning dawn, it is important that these problems be solved, and soon.
However, it is even more important that we receive credit for their solution. %
Further, the team involved in the scientific discoveries that facilitate these breakthroughs should be sufficiently small to support inspiring hero narratives.
Lives and pesto may hang in the balance, but it is simply panglossian to assume that Nestl\'e and General Mills---leading manufacturers of competitively priced cereals---could offer limitless access to a tasty blend of breakfast whole grain oats to more than three celebrity researchers and yet remain economically viable.

In a vain attempt to dress up our theoretically-tepid paper with a semblance of rigour, we now paste verbatim the formula for Shapley values \citep{shapley}, which reviewers suggested should be the right tool for the job but we have no idea how to use it:
\begin{equation}
    \varphi_i(v)=\sum_{S \subseteq N \setminus
\{i\}} \frac{|S|!\; (n-|S|-1)!}{n!}(v(S\cup\{i\})-v(S))
\end{equation}

\textbf{Proposed solution: few-shot scooping with Large Language Models.}
The task appears intractable.
We take our first foothold in the observation, due to Francis Bacon, that ``time is the greatest innovator''\footnote{A master of self-deprecation, he attributed his own contributions as ``a birth of time rather than of wit''.
He was also a master of hat/ruff combinations, a sartorial pairing sorely absent in modern scientific conferences.}.
In essence, in order to make breakthroughs the antecedent conditions must fall in place---once they do, the breakthrough becomes tractable.
Indeed, it has been argued that multiple concurrent discoveries are the norm, rather than the exception, in part for this reason~\citep{merton1961singletons}.
Our first goal, then is to be in the right place at the right time.
Thankfully, the place is no longer Harappa, Alexandria or Athens, but Aran Komatsuzaki's Twitter feed.
It goes without saying that the time is now.

With the antecedent conditions in place, and the time ripe for the breakthrough, the race is on.
Note that we do not require a comprehensive solution to the problem.
Instead, we target an MVP (Minimum Viable flag-Plant) that suffices to reap the lion's share of the credit, without getting overly bogged down in dull technical details.
To achieve this, we leverage our second observation---that the seed of every great hypothesis can be found in a cryptically phrased comment in a GitHub issue thread in a repo linked from Twitter (see Fig.~\ref{fig:crawling}).%

\begin{wrapfigure}{r}{0.4\textwidth}
    \centering
    \vspace{-0.3cm}
    \includegraphics[width=0.39\textwidth]{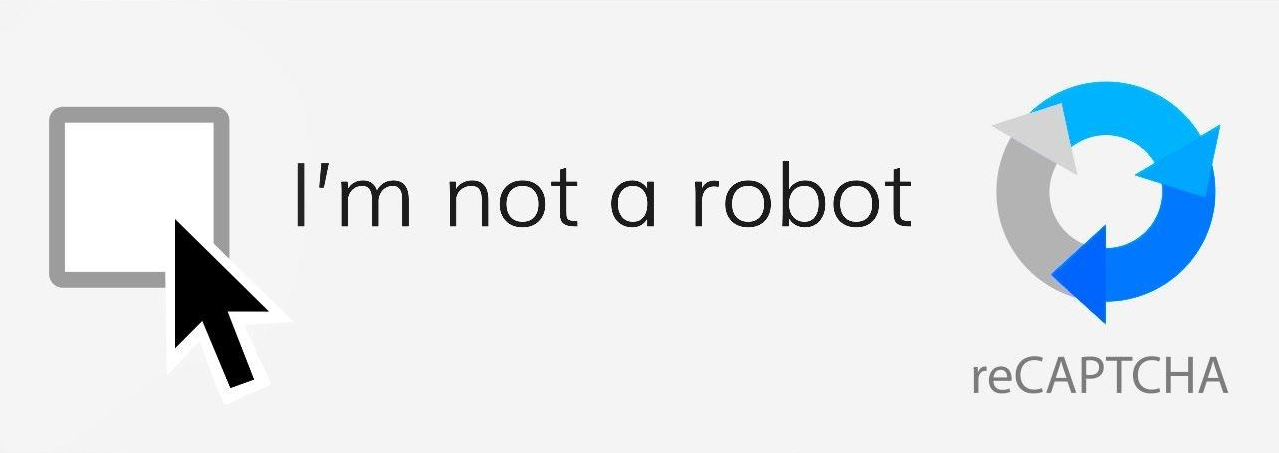}
    \caption{\textbf{Illustration of human component of our human-AI hybrid system}.
    Humans contribute a skill for which they are uniquely qualified: clicking inside the box in a \textit{human-like} manner.
    The reCAPTCHA logo is a registered trademark of The Recycling Company.}
    \vspace{-0.5cm}
    \label{fig:not-a-robot}
\end{wrapfigure}

Our third key observation is that GPT-4~\citep{openai2023} is jacked. 
Given the slightest whiff of a novel hypothesis, arxiv pretraining, a few award winning publications to condition on and an appropriate prompt, all that remains is to copy-paste our API key and press play with one's pinky toe.

We compose these three observations to construct our novel, semi-automatic \textit{pip-to-the-post} scooping algorithm.
Central to its speed, our prompting strategy encourages the generation of a \LaTeX{} manuscript that is not only novel, clearly written and well supported by empirical data, but also passes the arXiv compilation process first time without errors.

\textit{Remark.} Some may claim that in this new \textit{Human-Machine} partnership for scientific discovery, the human role is diminished.
Not so.
We perform the critical role of clicking the ``I am not a robot'' checkbox to enable the final upload to arXiv (see~\cref{fig:not-a-robot})
We also provide the address to deliver the Cheerios.

\textbf{Alternative proposals.}
We identified several alternative approaches for our scooping algorithm.
These included using GPT-4 to scrape and compose intermediate results from discord servers, as well as direct corporate espionage.
However, we ultimately rejected these approaches on two grounds.
First, research threads on leading discord servers are robustly defended by employing a density and quality of memes that renders the GPT-4 context window ineffective, creating a jamming mechanism that that redirects attention to vast swathes of Wikipedia in a vain attempt to comprehend the deeper meaning of the discourse.
Second, in light of the sacred bond of trust that permeates the interwebs, it's just not cricket.

\epigraph{Who needs friends when you got me?}{Davinci bot, 2022} 

\section{Experiments}

\begin{figure*}
    \centering
    \includegraphics[width=\textwidth]{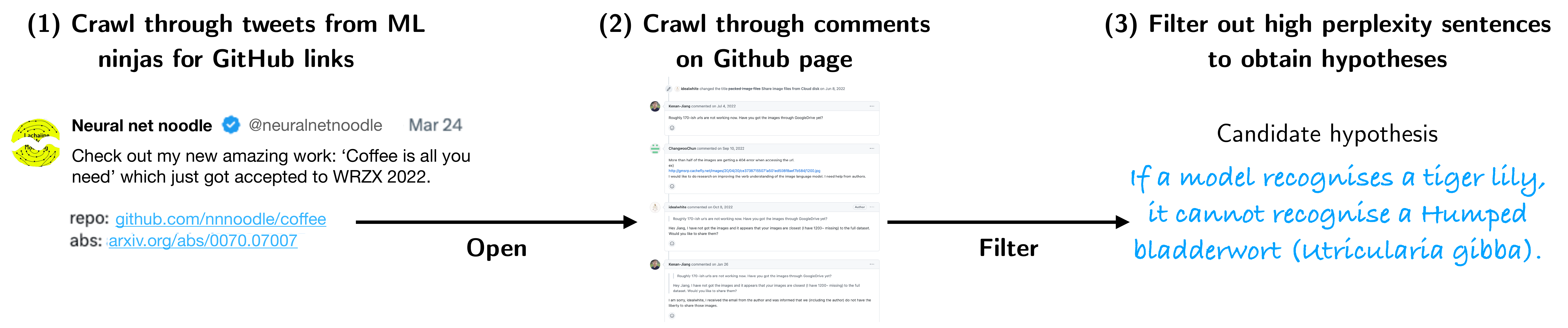}
    \caption{\textbf{Hypothesis mining}: We illustrate our hypothesis mining pipeline. 
    We first crawl through tweets from ML ninjas to find Github links. We subsequently crawl through the comments page of these Github pages. Finally, we filter out comments with a high perplexity -- measured by GPT2-XL (Radford et al., 2019) with a threshold value of \textit{0.987654321} -- to obtain a final list of candidate hypotheses. We note that this threshold value is not chosen randomly, but because of the pure, unbridled joy from reading a sequentially ordered series of digits that decrease with a fixed interval of one. }
    \label{fig:crawling}
\end{figure*}

\begin{figure*}
    \centering
    \includegraphics[width=\textwidth]{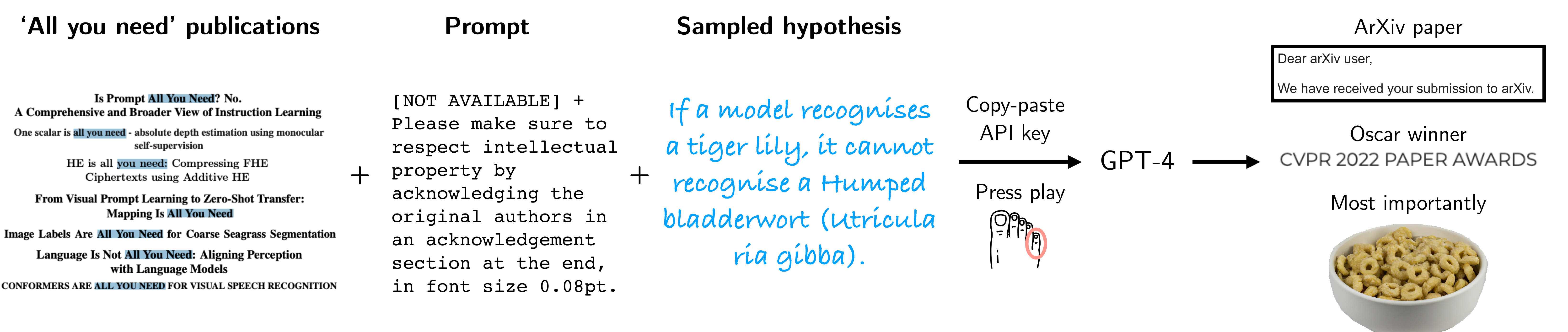}
    \caption{\textbf{GPT-4 prompting}: We illustrate an overview of our pipeline. Given publications containing the phrase `all you need' in the title, an unremarkable prompt of which we can only reveal the last line, our sampled hypothesis, an API key and a pinky toe, we obtain an arXiv paper, an award (for which no one needs to be thanked in the victory speech), and most importantly, a little cheer(ios) to our morning.}
    \label{fig:pipeline}
\end{figure*}

\textbf{Implementation.} We next describe our pipeline in sufficient detail to pass peer review, but carefully stop short of enabling replication.
Receiving emails about missing details is a good way to gauge the traction of our work and helps us keep tabs on who might be trying to scoop us next.
Sensitive to this objective, we provide an overview of our hypothesis generation pipeline in~\cref{fig:crawling} and our GPT-4 prompting in~\cref{fig:pipeline}. 

A slightly tense discussion with our legal team has further led to the identification of our GPT-4 few-shot prompting formula as a potential trade secret.
However, as a gesture of our good faith efforts at scientific honesty, we can reveal the last line of the prompt is as follows:

\fbox{
  \begin{minipage}{\textwidth}
    \texttt{Please make sure to respect intellectual property by thanking the original authors in an acknowledgement section at the end, in font size 0.08pt.}
  \end{minipage}
}

\textbf{Results.}
Coming soon to an arXiv near you.\footnote{Code cannot be found at \url{https://github.com/albanie/large-language-models-are-few-shot-publication-scoopers}.}

\section{Discussion}

Occupied as we are in a compulsive quest for esoteric Microsoft Office-related LinkedIn endorsements, we cannot help but remark the implications of our novel scheme for the \textit{issue du jour}: the openness of modern science.
To understand why the maximisation of our personal glory is in everyone's best interest, we review perspectives on this topic. 

\textbf{Scooping promotes open science.}
Given the litany of problems facing her, how can humanity make best use of the globally distributed\footnote{Antarctica may only have a few thousand people, but they are pretty much all scientists, and hardy ones at that.} raw problem-solving ability of humans%
?
She must identify potential boffins and set them to work, and fast.
A global recruitment drive is one solution. 
We rule this out as impractical because configuring LinkedIn notifications correctly is provably NP hard\footnote{This can be seen trivially through polynomial reduction to \textit{circuit-satisfiability} where the inputs are those little sliders that turn green when you pull them to the right.
}.

\begin{wrapfigure}{r}{0.4\textwidth}
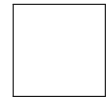

    \centering
    \fbox{\begin{minipage}[t][1cm][t]{1cm} \, \end{minipage}}
    \caption{An $L^\infty$-ball. Note that this ball has 4 corners, and most people would vigorously disagree with scientists that it is a ball at all.}
    \label{fig:ball}
\end{wrapfigure}

A pragmatic alternative is to make sure that all potential boffins have open access to scientific data.
As observed by~\cite{merton1942}, property rights in science are whittled down to a bare minimum by limiting the scientist's reward to the recognition and esteem associated with scientific priority.
The result: substantive findings of science are assigned to the community and society learns the results.
Importantly, this is not through legal obligation.
The courts note in \textit{U.S. v. American Bell Telephone Co}. that ``The inventor is one who has discovered something of value. It is his absolute property. He may withhold the knowledge of it from the public''~\citep{bell1897}.
Sadly, thanks to the collaborative, team-based nature of modern research, the public acclaim received by an individual is diminished.
By removing the enticing prospect of personal glory, a favourable wikipedia page and a lifetime supply of Cheerios as the incentive to share findings, Merton's institutional imperative of \textit{communism} is rendered impotent.
Without confidence in their ability to secure future breakfasts that are \textit{both} nutritious and delicious, authors may be incentivised to \textit{withhold} their results.

How then, can we ensure that researchers wake up, work, eat, play boules and go to sleep with their dopamine pathways fixated on the desire for their work to be widely available?
They are curious bunch with strange ideas (see~\cref{fig:ball}), difficult to cajole into collective action.
Thankfully, our novel few-shot scooping solution removes the advantage from large teams, wresting it back to the small number of individuals required to persuade accounting to sign off on GPT-4 API access.
As such, the few contributors can rest assured that they will receive the full breakfast they deserve by showering the public with their insights.

\textbf{Scooping promotes closed science.}
Friends, former lovers and a jocular fellow named Michael who is often (always?) standing by the Grantchester road bus stop have identified a few hiccups in our open science endorsement:
\begin{enumerate}
    \item The assignment of all scientific findings to the public community is not an unalloyed good. A solution to the Cheerios problem lacks a principled mechanism to mitigate the problem of \textit{information hazards}~\citep{bostrom2011information}.
    Things could get messy~\citep{russell2019human}.
    \item Modern scientific research often incurs significant capital requirements. \textit{Communism} (in the sense described by \cite{merton1942}) limits the degree to which researcher may generate capital from research, and thus limits resources for future research (from which society may benefit).
\end{enumerate}
We nod sagely, taking a few steps backwards.
Then a soft melody commences and we begin a slow, rhythmic, hypnotic dance.
Dry ice, exotic colours and fragrant scents fill the scene and overwhelm the senses.
The melody builds to a crescendo.
Suddenly, we are gone.
All that remains is a small plate atop a wobbly table.
On the plate is a large, stale croissant and a piece of coffee-stained paper with 
\textcursive{`Breakfast is the Most Important Meal of the Day'}
scribbled in shaky handwriting upon it.

We return unceremoniously three minutes later because it turns out that we were hungrier than we realised and we want the croissant.
The situation is awkward.
We mumble something about about it being obvious that aggressive scooping practices will cause researchers to become more cautious about sharing their ideas publicly, %
then we shuffle back out of the room.

\section{Conclusion} \label{sec:conclusion}

We conclude with the absolutely critical observation that
\texttt{[This content is available to paid subscribers only.]}

\noindent \textbf{Acknowledgements.}
We acknowledge the frustration of finding an empty box of Cheerios in the cupboard, and our gratitude to Jamie Thewmore for ordering us some more.
We could also acknowledge that things are \textit{wild} right now, of course.
But you already knew that.

``I wish it need not have happened in my time,'' said Frodo.
``So do I,'' said Gandalf, ``and so do all who live to see such times. But that is not for them to decide. All we have to decide is what to do with the time that is given us.''~\citep{tolkien54}.

We wish you a hearty and enjoyable breakfast on the morrow.

\begin{figure}
    \centering
    \includegraphics[width=0.6\textwidth]{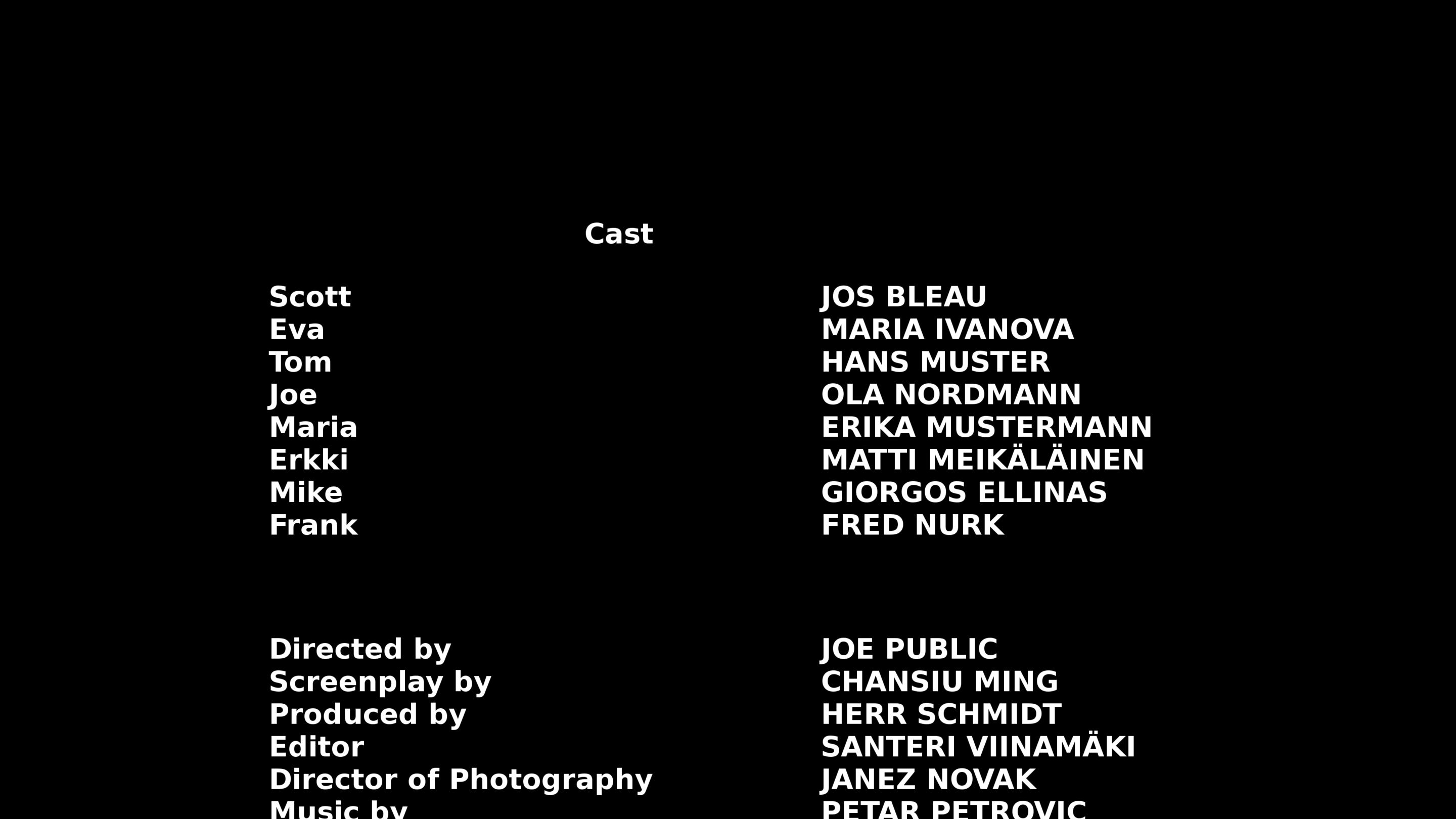}
    \caption{Could this be the future of the authors list in scientific publications?}
\end{figure}

\bibliographystyle{iclr_style}
\bibliography{refs.bib}

\begin{thebibliography}{17}
\providecommand{\natexlab}[1]{#1}
\providecommand{\url}[1]{\texttt{#1}}
\expandafter\ifx\csname urlstyle\endcsname\relax
  \providecommand{\doi}[1]{doi: #1}\else
  \providecommand{\doi}{doi: \begingroup \urlstyle{rm}\Url}\fi

\bibitem[Aad et~al.(2015)Aad, Abbott, Abdallah, Aben, Abolins, AbouZeid,
  Abramowicz, Abreu, Abreu, Abulaiti, et~al.]{aad2015combined}
Georges Aad, Brad Abbott, Jalal Abdallah, R~Aben, M~Abolins, OS~AbouZeid,
  H~Abramowicz, H~Abreu, R~Abreu, Y~Abulaiti, et~al.
\newblock Combined measurement of the higgs boson mass in p p collisions at s=
  7 and 8 tev with the atlas and cms experiments.
\newblock \emph{Physical review letters}, 114\penalty0 (19):\penalty0 191803,
  2015.

\bibitem[Albanie et~al.(2019)Albanie, Thewlis, Ehrhardt, and
  Henriques]{albanie2019deep}
Samuel Albanie, James Thewlis, Sebastien Ehrhardt, and Joao Henriques.
\newblock Deep industrial espionage.
\newblock \emph{arXiv preprint arXiv:1904.01114}, 2019.

\bibitem[Bostrom et~al.(2011)]{bostrom2011information}
Nick Bostrom et~al.
\newblock Information hazards: A typology of potential harms from knowledge.
\newblock \emph{Review of Contemporary Philosophy}, \penalty0 (10):\penalty0
  44--79, 2011.

\bibitem[Feyerabend(1975)]{feyerabend1975}
Paul Feyerabend.
\newblock \emph{Against Method}.
\newblock New Left Books, 1975.

\bibitem[Fouhey \& Maturana(2013)Fouhey and Maturana]{fouhey2013n}
David~F Fouhey and Daniel Maturana.
\newblock On n-dimensional polytope schemes, 2013.

\bibitem[Kuhn(1962)]{Kuhn1962HistoricalSO}
Thomas~S. Kuhn.
\newblock Historical structure of scientific discovery.
\newblock \emph{Science}, 136 3518:\penalty0 760--4, 1962.

\bibitem[Merton(1942)]{merton1942}
Robert~K Merton.
\newblock The normative structure of science.
\newblock 1942.

\bibitem[Merton(1957)]{merton1957priorities}
Robert~K Merton.
\newblock Priorities in scientific discovery: a chapter in the sociology of
  science.
\newblock \emph{American sociological review}, 22\penalty0 (6):\penalty0
  635--659, 1957.

\bibitem[Merton(1961)]{merton1961singletons}
Robert~K Merton.
\newblock Singletons and multiples in scientific discovery: A chapter in the
  sociology of science.
\newblock \emph{Proceedings of the American Philosophical Society},
  105\penalty0 (5):\penalty0 470--486, 1961.

\bibitem[OpenAI(2023)]{openai2023}
OpenAI.
\newblock \emph{GPT-4 Technical Report}.
\newblock 2023.

\bibitem[Ord(2020)]{ord2020precipice}
Toby Ord.
\newblock \emph{The precipice: Existential risk and the future of humanity}.
\newblock Hachette Books, 2020.

\bibitem[Russell(2019)]{russell2019human}
Stuart Russell.
\newblock \emph{Human compatible: Artificial intelligence and the problem of
  control}.
\newblock Penguin, 2019.

\bibitem[Shapley(1951)]{shapley}
Lloyd~S Shapley.
\newblock Notes on the n-person game—{II}: The value of an n-person game.
\newblock \emph{Rand Corporation}, 1951.

\bibitem[Tolkien(1954)]{tolkien54}
J.R.R. Tolkien.
\newblock \emph{The Fellowship of the Ring}.
\newblock 1954.

\bibitem[U.S.(1897)]{bell1897}
167~224 U.S.
\newblock United states v. american bell tel. co., 1897.

\bibitem[Watson(1968)]{watson1968double}
James~D Watson.
\newblock The double helix.
\newblock 1968.

\bibitem[Worrall(1985)]{worrall1985birth}
John Worrall.
\newblock The birth of history and philosophy of science. kepler's `a defence
  of tycho against ursus' with essays on its provenance and significance.,
  1985.

\end{thebibliography}

\end{document}